\documentclass[USenglish]{article}

\usepackage[utf8]{inputenc}
\usepackage{natbib}
\usepackage[small]{dgruyter}
\usepackage{microtype}
\setcitestyle{authoryear,open={(},close={)}}

\begin{document}

\articletype{Research note}

\runningauthor{Lopez}
\title{Bigger data, better questions, and a return to fourth down behavior: an introduction to a special issue on tracking data in the National football League}
  
\author*[1]{Michael J. Lopez}
\affil[1]{National Football League \& Skidmore College, Email: michael.lopez@nfl.com}
\runningtitle{NFL player tracking intro}
\subtitle{}
\abstract{Most historical National Football League (NFL) analysis, both mainstream and academic, has relied on public, play-level data to generate team and player comparisons. Given the number of oft omitted variables that impact on-field results, such as play call, game situation, and opponent strength, findings tend to be more anecdotal than actionable. With the release of player tracking data, however, analysts can better ask and answer questions to isolate skill and strategy. In this article, we highlight the limitations of traditional analyses, and use a decades-old punching bag for analysts, fourth-down strategy, as a microcosm for why tracking data is needed. Specifically, we assert that, in absence of using the precise yardage needed for a first down, past findings supporting an aggressive fourth down strategy may have been overstated. Next, we synthesize recent work that comprises this special Journal of Quantitative Analysis in Sports issue into player tracking data in football. Finally, we conclude with some best practices and limitations regarding usage of this data. The release of player tracking data marks a transition for the league and its’ analysts, and we hope this issue helps guide innovation in football analytics for years to come.}
  \keywords{National Football League; player tracking; causal inference; fourth downs}
  \startpage{1}
  \aop

\maketitle

\section{The problem with football data} 

Across the physical and social sciences, the gold standard for identifying causes and effects of certain behaviors, therapies, or interventions is the randomized experiment. Randomization is attractive because subjects that receive one treatment are, in expectation, comparable to those that receive another treatment. When examining an outcome of interest in a randomized trial, one can be confident that there are no underlying and unmeasurable differences (e.g, confounding variables) that would be responsible for causing said results. Succinctly – randomized designs ensure that apples are compared to apples. 

Analyzing football data is like that, except the opposite. That is, all virtually all football data is observational, which means that any study of athlete or team behavior is potentially confounded by other variables linked to the game or player. Because of this, it is often quite difficult to rule out whether extraneous factors related to players and games are responsible for findings. In football, we rarely get to compare apples to apples.

Consider the well-established argument in NFL analytics circles that teams should pass more and run less \citep{kovash2009professionals, fivethirtyeightpass}. Unlike what would happen in randomized designs, teams don’t flip a coin to decide if they’ll attempt a pass. Instead, coaches call a pass play based on, among other factors, quarterback skill, game situation, and opposing defense, many of which are traits that likely differ from settings that call for run plays. Certain drivers of play choice, including down, distance needed for a first down, and score differential, are known prior to a play. But several other variables, such as quarterback health, defensive formation and personnel, and pre-snap movement, are both (i) likewise linked to whether or not a team attempts a pass, and (ii) difficult to quantify.

Not only are there several factors that go into NFL play-calling, but many cannot be measured using traditional data. An artifact of the NFL's historical data collection is that the hundreds of player movements and decisions in a play are reduced to one observation, one row in a data set. Worse, at least in terms of public data, most of the 22 players on the field at a given time aren't even recorded as being there \citep{schatz2005football}. At around 160 plays per game, it is feasible to look at play-level outcomes such as win probability, expected points, run/pass strategy, and fourth down behavior, while conditioning on what we know about plays before they happen. But both before and within a play, events in football are reactionary. Substitutions by one team lead to changes in personnel by the other. Formations by the defense lead to audibles by the offense. Motion from the slot back spurs a new defensive coverage scheme. And even after the play, hip placement of defensive backs creates new cuts from wide receivers and a lineman's first step signals an opposing linebacker's read. Under this complex interplay, nothing is as simple as whether or not a team's coach called a passing play, which makes trying to untangle the marginal effect of passing versus running exceedingly difficult. Thus, even when conditioning on several variables, such as \cite{fivethirtyeightpass}, which includes number of defenders near the line-of-scrimmage and other play and game-level characteristics, we are likely not accounting for something else. 

Why do these differences matter? The answer lies in how we leverage analytical insight to impact the game. It means one thing to find that passing results in better outcomes (yards per play, expected yards per play, win probability added) than run plays. Analysis is often improved by conditioning on traits such as down, distance, and formation. But it means substantially more to find that among teams that ran the ball, they’d have been better off had they passed more often. This is a tricky but important distinction. This second claim is a causal one, and cannot be made with certitude unless we are able to condition on all relevant football variables. Researchers are unable to condition on variables we do not have, which jeopardizes any attempt to establish causality in football. 

Beginning in 2016, those 160 observations per game turned into roughly 300,000. That season, each player was equipped with radio-frequency identification (RFID) chips in each of his shoulder pads, observations that provide the location of each player and the football at roughly 10 frames-per-second, wherever he goes on the field. From player movement, it is straightforward to calculate speed, angle, and acceleration. These data are termed the NFL's "Next Gen Stats" (see $https://nextgenstats.nfl.com/$ for more insight and summary metrics). For the first time, analysts can dig beyond play-level analyses to better understand the game. Variables that used to only show up in scouting reports -- ones such as route type, running back space, or defensive back coverage -- can now be estimated directly from data. Soon, analysts will be breaking down a game before their coaches have even watched the film. 

Given the novelty of player tracking data, much remains unknown about the questions that can be answered from it. But for statisticians, the goal should be clear – how can we leverage this data to ensure we are comparing apples to apples? 

\section{The importance of unmeasured confounding: a unique example}

Over the last several years, the lowest hanging fruit among NFL analytics enthusiasts has been team behavior on fourth down. Authors as far back as \cite{carter1978note}, and including more recent work from \cite{romer2006firms}, \cite{NYT4Bot}, and the author of this manuscript (\cite{yamlost}), have argued that teams are too passive on fourth down. Roughly, it has been argued that a more aggressive strategy is worth 0.4 wins per year \citep{romer2006firms, yamlost}. In a limited 16-game season, that fractional gain takes on an added importance.

Traditional fourth-down analyses has compared play outcomes (such as success rates, expected points, or win probability) before and after potential conversion attempts. The crux of these works requires some level of extrapolation regarding the outcomes for teams that did not go for it, if they were instead to have gone for it. For example, \cite{romer2006firms} used success rate, \cite{NYT4Botconstruction} used expected points, and \cite{yamlost} used win probability, all to imply that teams that did not go for it would have been better off going for it. 

Each fourth-down analysis framework mentioned above assumed some level of equipoise between teams that did not go for it and those that did. \cite{yamlost} went as far as using propensity-score based matching techniques to ensure that teams that did not go for it were compared to similar teams that did go for it. But even the most novel of matching techniques designed for observational data cannot get around the limitation of unmeasured confounding. That is, what if there was an unmeasured characteristic of teams that went for it that fundamentally differed from those that did not go for it, one that simultaneously impacted both coaches’ decisions and play outcomes?

Turns out, there is one – the precise distance needed for a first down.

\subsection{How precise distance impacts fourth-down strategy}

In the gathering and disseminating of play-by-play data, NFL game-day assistants assign an integer value to each distance needed for a first down. All distances between 0 and 2 yards (not inclusive) are supposed to be listed as 4th-and-1’s, with successive buckets consisting of one-yard intervals. So, both 4th-and-0.1 inches and 4th-and-71.9 inches are listed as 4th-and-1’s. Likewise, distances of 4th-and-72 inches to 4th-and-107.9 inches are intended to be listed as 4th-and-2's.  

\begin{figure}
    \centering
    \includegraphics[width = \textwidth]{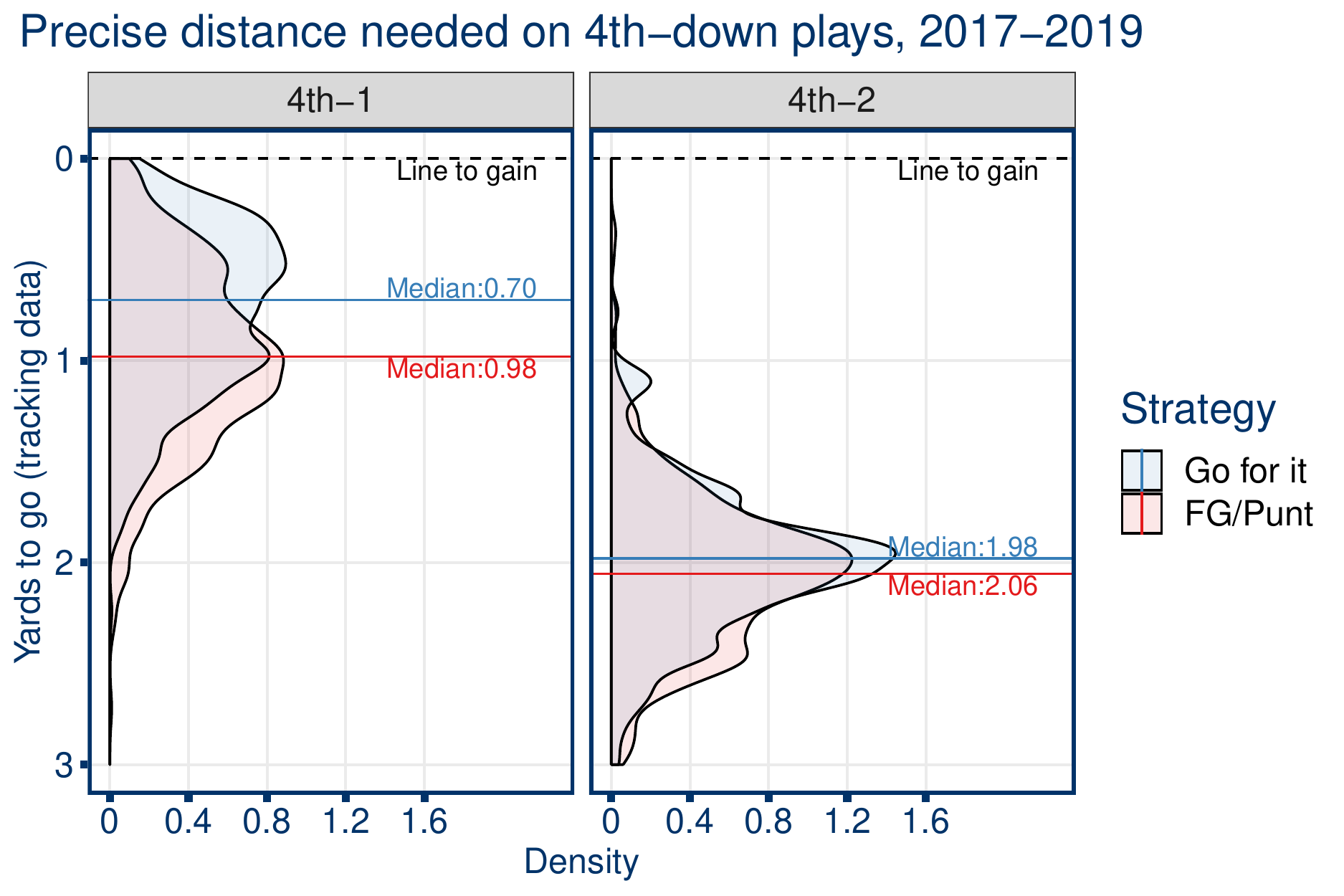}
    \caption{Exact distances needed for a first down, split by team decision (Go for it, FG/Punt) and NFL play-by-play yard line categorization (4th-and-1, 4th-and-2). Teams that went for it were closer to the line to gain.}
    \label{fig:my_label1}
\end{figure}
 
Figure \ref{fig:my_label1} shows two pairs of density plots showing the distributions of precise yards needed for a first down. Precise yardage was identified by comparing the football location on fourth down to the line to gain, with the latter obtained by using the football location on first down.\footnote{On goal-to-go plays, the goal line is used as the line to gain.} The chart is split by the integer categorization in the NFL play-by-play data (either a 4th-and-1 or a 4th-and-2). Fourth down plays from the 2017 through 2019 regular seasons are used.

Teams that went for it on 4th-and-1 were typically 0.70 yards away from the line to gain; teams that did not go for it, meanwhile, were 0.98 yards away. On 4th-and-2 plays, teams that went for it did so from a median distance of 1.98 yards away, compared to 2.06 yards away for teams that did not go for it.\footnote{We also used a second data wrangling strategy, where we compared the distance on the RFID chip embedded in the football to the RFID chip in the sideline chain that demarks the first down line. Differences in the numbers provided were negligible.}   

\begin{figure}
    \centering
    \includegraphics[width = \textwidth]{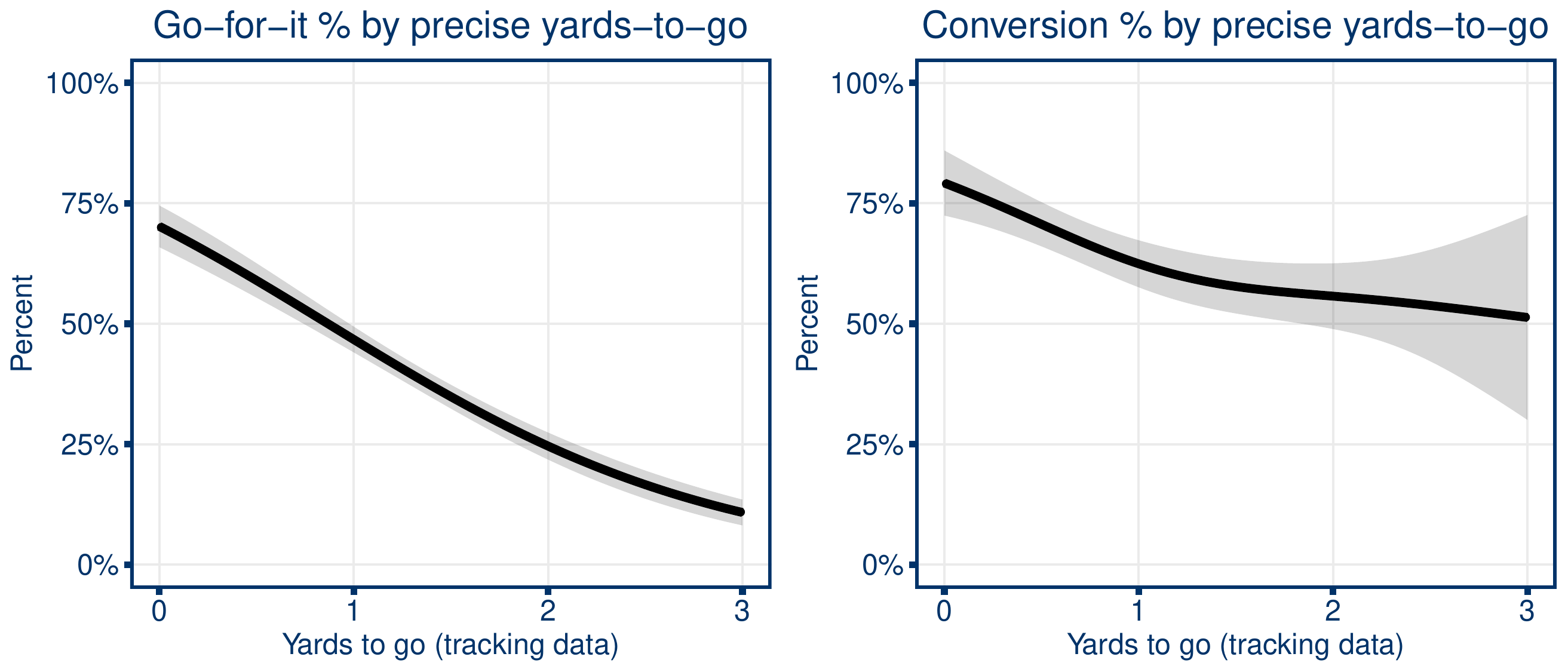}
    \caption{Likelihood of going for it (left side) on 4th-down, and likelihood of a successful conversion (right side) on 4th-down. Each line shows the estimates from a generalized additive model of outcome (either going for it, or of successfully going for it) given the precise distance needed for a first down. Teams with shorter distances are more likely to go for it and to convert.} 
    \label{fig:my_label2}
\end{figure}

The precise distance needed for a first down impacts both the attempt rate (among all teams) and the success rate (among teams that went for it), as shown in Figure \ref{fig:my_label2}. The left side of Figure \ref{fig:my_label2} provides estimates from a generalized additive model (GAM) of attempt rate (Did teams go for it (Y/N), conditional on the precise distance needed for a first down). Teams facing 4th-and-inches went for it about 70\% of the time, while teams in *long* 4th-and-1 situations went for it about 30\% of the time. The right side of Figure \ref{fig:my_label2} highlights how the rate of success varied based on precise distance needed for a first down, using estimates from a separate GAM (Did teams gain a first down (Y/N), conditional on the precise distance needed). On 4th-and-inches, teams converted roughly 79\% of the time, while in *long* 4th-and-1 situations, they converted around 55\% of the time.

What does this imply? 

Because they had further to go for a first down, teams that did not go for it did not have the same chances for success as the teams that did go for it. Thus, findings that inferred success rates, expected points, or win probability outcomes from teams that went for it almost assuredly overestimated the benefit of going for it on fourth down. For years, we failed to compare apples to apples.

Interestingly, although the GAM's in Figure \ref{fig:my_label2} allow for a non-linear relationship between distance and each outcome, each curve looks somewhat linear. If we instead fit a line using the integer categorization of distance, we would see roughly the same figure. That is, even though we were able to use more precise data, our results still matched how we could have predicted coaches to behave using only static data. 

Replicating the approach of \cite{yamlost} can help approximate what this means as far as the value of 4th-down aggressiveness. While those authors used data from prior to the 2016 season, we use the 2017 through 2019 seasons only, with an identical approach and code provided at $https://github.com/statsbylopez/nfl-fourth-down/tree/master/Code$. We replicate under two scenarios. First, we use the play-by-play distance category to check results over the past three seasons. Second, we use the more precise distance that the offense needed in place of the integer distance. As in \cite{yamlost}, we only use plays in the ``go for it'' range of the 4th-down-bot \citep{NYT4Bot}.

Using the play-by-play distance category (4th-and-1, 4th-and-2, etc), we find that an aggressive fourth-down strategy would have been worth, in expectation, an extra 0.35 wins per-team per-year from 2017 to 2019. Among teams that did not go for, we find an estimated 3.8\% difference in win probability added per-play had they instead gone for it (95\% CI, 2.6\% to 4.9\%). This is somewhat in line with the original findings, with a slight drop perhaps driven by recent upticks in team aggressiveness on fourth down \citep{slate_4th}. 

When accounting for the precise distance needed for a first down (instead of the integer distance), the benefit of an aggressive fourth-down strategy drops to an extra 0.22 wins per year. Alternatively, the average difference in win probability added per-play is estimated at 2.2\% (95\% CI, 0.6\% to 3.4\%). For both per-play win probability and per-team benefit in terms of wins-added per-year, roughly 40\% of the benefit to an aggressive fourth down strategy is negated when accounting for actual distance needed for a first down, a previously unmeasured variable. 

Limitations in standard fourth down analysis are further broadened in Figure \ref{fig:my_label3}. In Figure \ref{fig:my_label3}, the precise distance on each 4th-and-1 play from tracking data is shown on the y-axis, relative to the estimated probability that the offense went for it (x-axis). This probability is calculated using play-by-play data only (we use the \cite{yamlost} model that includes 17 variables such as yard line, time, score, timeouts remaining, and pre-play win probability). Two smoothed trend curves are shown in Figure \ref{fig:my_label3} -- one apiece for teams that went for it (in blue) and for teams that did not go for it (red). Across nearly the entirety of go-for-it probability, the differences in precise distance needed for a first down between the two curves are larger than the aggregated difference (0.28 yards) shown in Figure \ref{fig:my_label1}, a result akin to Simpson's Paradox. For example, among teams with a low probability of going for it, teams that went for it were more than half a yard closer to the line to gain relative to teams that did not go for it. In other words, by conditioning on play-by-play data, the differences in tracking data distance between teams that went for it and those that did not go for it actually grew more extreme. From a causal inference perspective, by matching on observed covariates but not an unobserved one, \cite{yamlost} exacerbated the imbalance in the precise distance teams needed for a first down, potentially worsening the bias in the estimated effect. This phenomenon is known as squeezing the balloon \citep{brooks2013squeezing}.

\begin{figure}
    \centering
    \includegraphics[width = \textwidth]{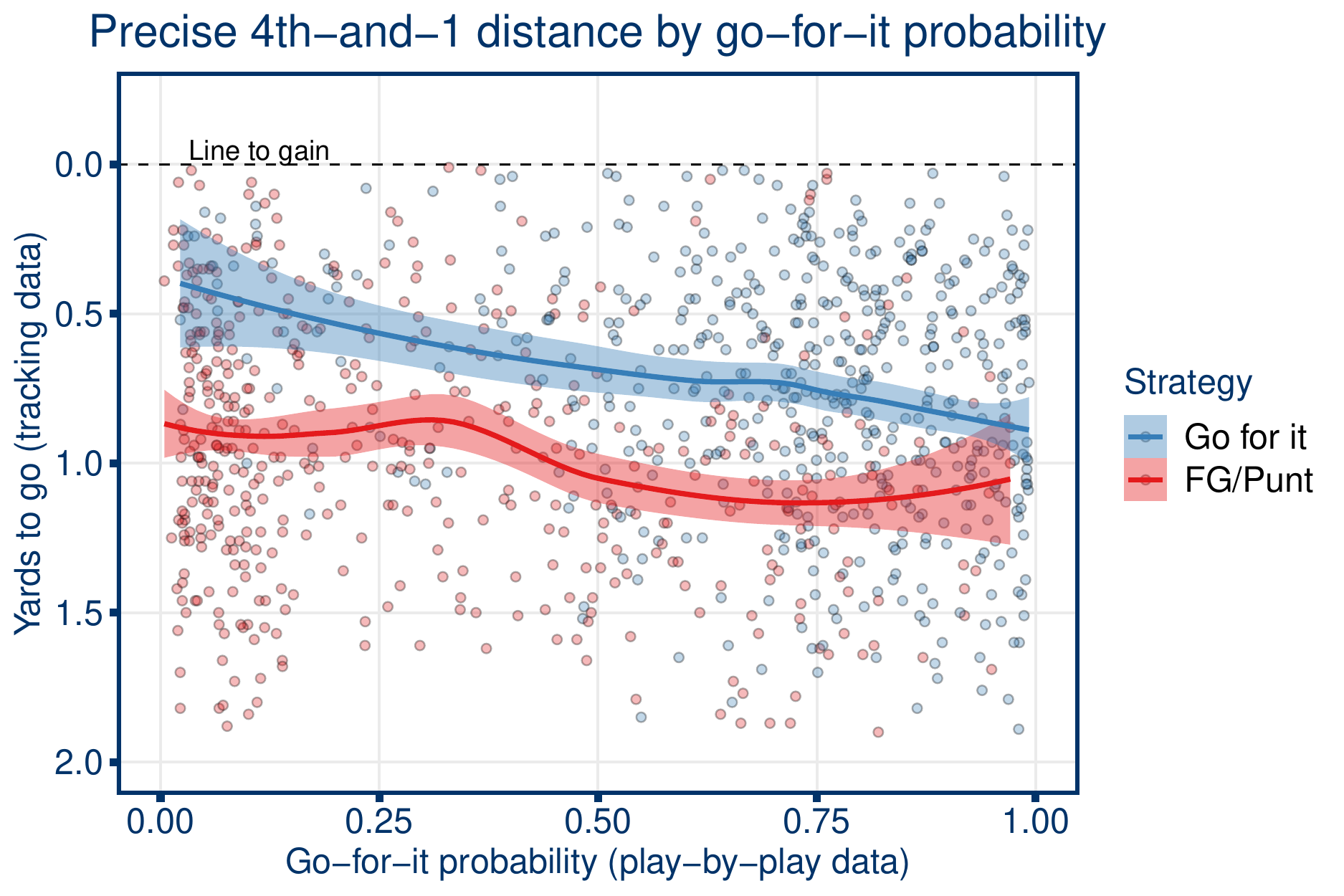}
    \caption{Exact distance (using tracking data) needed for a first down on 4th-and-1 plays, split by team decision (Go for it, FG/Punt) and compared to go-for-it probability (estimated using play-by-play data). Teams that went for it were closer to the line to gain for all levels of go-for-it probability. This chart highlights that when conditioning on observed variables (in this example, using play-by-play data), differences in previously unobserved variables (in this example, tracking data distance) can grow more extreme. In Figure \ref{fig:my_label1}, the difference in distance needed for a first down on 4th-and-1 plays was 0.28 yards, which is less than the difference between the blue and red curves above.}
    \label{fig:my_label3}
\end{figure}

Though an attenuation of the effect size matches our intuition that going for it on fourth down is generally less advisable from longer distances, the primary lesson here extends beyond 4th-down strategy. Indeed, there could be other reasons why the benefit of aggressive approach remains underestimated \citep{romer2006firms}. Instead, we aim to highlight the surprising insight that tracking data can provide. Repeated analyses of something in the game of football, in this case, fourth downs, had told us that coaches should ``stop being stupid'' with how they were acting \citep{apex}. But there was more to the story, in this case something intrinsically different about teams that went for it and teams that did not, which was responsible for at least part of the original findings. 

Indeed, in football, there's almost always more to the story. 

\section{Crowdsourcing insight into player tracking data}

Each of the six papers included in this special issue on player tracking data in the National Football League use data from the Big Data Bowl, an event originating from the NFL league office designed to crowdsource public insight into tracking data, inspire analytically-inclined fans, curate ideas for team staffers, and spur data-driven innovation in football. The homepage for the 2019 event can be found at \url{https://operations.nfl.com/the-game/big-data-bowl/2019-big-data-bowl/}. Big Data Bowl participants were given full, raw player tracking data from the first six weeks of the 2017 regular season, as well as game, play, and player characteristics. More than 1800 participants signed up for the contest.

The papers in this special issue each highlight the multitude of ways in which football analysts can better understand movement, tendencies, and the spatial constructs underpinning football. Moreover, and in linking to the introductory fourth down example, tracking data has allowed each of the authors to make more apples-to-apples comparisons than is possible with play-level information. 

We'll start with two papers on the principal movement of NFL wide receivers, a passing route. Chu, Reyers, Thomson, and Wu -- winners in the College section of the 2019 Big Data Bowl -- derive passing routes using model-based clustering, while Kinney matches receiver movement to the traditional NFL passing tree. In addition to making route labeling instantaneous (which would save coaches a tremendous amount of time), the intention behind both Chu et al and Kinney is to find similarity in movement. There's insight in both comparing one route to another (e.g, which routes create the most separation from defenders), but also in looking within routes themselves (e.g, which player breaks the quickest on a comeback, or how fast to players get downfield on a post). 

Ventura and Dutta likewise look for similarity on passing plays, but instead focus on defensive coverages schemes. As in the wide receiver examples, common movement patterns emerge from the tracking data, representing cornerback coverage in each of zone and man defensive schemes. Defensive tendencies also play a role in Mallepalle, Yurko, Pelechrinis, and Ventura's paper that both (i) provides an approach for extracting raw NGS data from images and (ii) contrasts completion percentages for quarterbacks and defenses across field locations.

The final pair of papers take a more omniprescent approach to the entirety of a football play. First, Evans and Deshpande consider unobserved passing outcomes via Bayesian Additive Regression Trees. By being able to estimate receiver catch probabilities across an entire route, Deshpande and Evans open the door for identifying which coach called the best play, which quarterback made the best decision, or which wide receiver was most easily able to get open. Finally, Yurko, Matano, Richardson, Granered, Pospisil, Pelechrinis, and Ventura look frame by frame within a play to better understand the value of each player movement. Long short-term memory neural networks, in combination with conditional density estimation, allow for real-time estimates of, as an example, where a running back is likely to be tackled.

Figure 10 of the Yurko et al paper perhaps best exemplifies the value of tracking data in football for analyzing player ability. In this example, ball carrier Cordarelle Patterson received the handoff near midfield on a 2nd-and-short. By using the speed and movements of players on the field, Yurko et al estimate an expectation that Patterson would gain 15 yards. When cross referencing play-by-play data, however, only about 5\% of 2nd-and-short running plays near midfield are that successful. That is, before Patterson has even made a move with the ball, we know better than to judge his success against similar 2nd-and-short plays, and instead can focus on other examples where ball carriers had such high expectations.

\section{Conclusion and next steps}

This manuscript highlights several use cases of NFL tracking data, including both old and new research questions. In particular, we return to one of the league's oldest findings -- that teams should be more aggressive on fourth down -- to suggest that previous work may have overestimated the effect of going for it. Specifically, given the precise distance needed for a first down, teams that went for it on fourth down tended to do so from shorter distances, even when conditioning on the play-by-play yardage category. More generally, we summarize how the articles in this JQAS special issue on player tracking data will help shape the future of NFL analytics work. 

Although we explicate on the value of player tracking insight in the NFL, it is important to acknowledge that this data is not a panacea for all football problems. Given the complexity of the game, there will always be fundamental football questions that data alone cannot precisely answer. Additionally, player tracking data is more arduous to analyze when compared to traditional play-by-play data; anecdotally, nearly every entrant to the league's Big Data Bowl wished they could have had more time to refine their work. To wit, here is a list of best practices and caveats for working with tracking data in the NFL. 

1. Tracking data contains the $x$ and $y$ coordinates for each player and the football, collected at roughly 10 frames-per-second. Locational information is provided by signals sent from radio-frequency identification (RFID) chips that are placed inside each player's shoulder pads and inside the football. Speed, orientation, and distance traveled are straightforward to calculate using the tracking information, and are provided by the NFL's Next Gen Stats group. The typical game can contain anywhere from 250,000 to 350,000 rows of data (1 row for each player on the field, on each play, at each time stamp) on which actual game action is occurring. Players are also tracked before and after plays (this information was not provided as part of the Big Data Bowl, and is generally seen as less pertinent). The $z$ coordinate is not measured (e.g, height of the player, or height of the ball), nor can the precise location of helmets, arms, and legs be verified or easily estimated.

2. The field coordinates are fixed at each NFL stadia, as shown in Figure \ref{fig:my_label3}. From left to right, the length of the field spans from $x$ = 0 to $x$ = 120 (units are in yards), while the width of the field spans $y$ = 0 to $y$ = 160/3. Often, the first step in any analysis of tracking data is to ensure offensive teams are moving in the same direction. This requires flipping roughly half of a game's offensive plays from one direction to the other, while creating new $x$ (subtracted from 120) and $y$ (subtracted from 160/3) coordinates. Additionally, standardizing by the play's line-of-scrimmage may be warranted. 

3. Several play-specific traits remain unknown even when looking at tracking data. These include the initial play call, if the quarterback or coach called an audible, how the defense would have lined up if the offense used a different formation, if a wide receiver ran the correct route, etc. Each of these variables may be pertinent to more precisely estimate, as an example, the value of passing versus running. The absence of important play-level qualities highlights the need for analysts to work directly and cohesively with football experts in order to maximize the value of tracking data. 

4. Some specifics about the data that researchers may want to be aware of. First, given updates to the RFID tags prior to the start of each season, small differences in speed measurements may exist from one year to the next. Additionally, the coordinates on the football are considered to be slightly less reliable than the coordinates on the players. Next, while analyzing maximum speed for players is often an easy-to-understand step, researchers should be wary that occasionally this maximum speed is reached while (or immediately after) a player is hit or tackled by an opponent. That said, tracking data is considered quite dependable; according to the Next Gen Stats group, location information is accurate to within +/- 12 inches, and reliable data has been collected on 99.999\% of the entirety of players and games over the last three seasons.

\begin{figure}
    \centering
    \includegraphics[width = \textwidth]{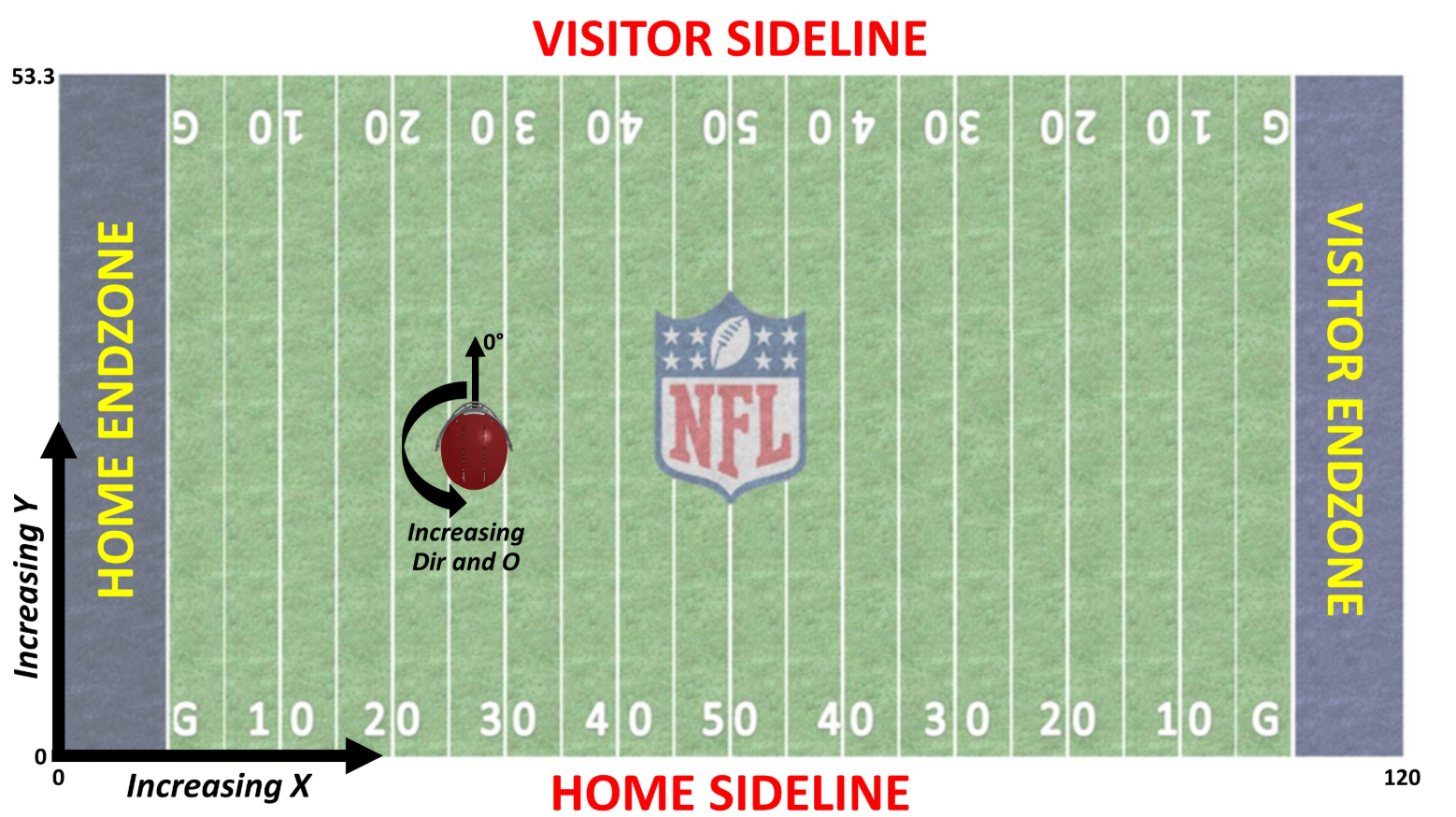}
    \caption{Coordinates for working with player tracking data. Each stadia is equipped such that the home and visiting team end zones are fixed throughout the game. However, the end zones that teams defend in a game are not determined until the start of each half, and those directions change at the conclusion of the first and third quarters.} 
    \label{fig:my_label3}
\end{figure}

\textbf{}

For years, data-driven innovation in the football was limited, and the NFL was, rightly or not, perceived to be trailing other leagues in terms of how teams used analytics. But insight lagged, in part, because so too did data. Behind player tracking insight, such excuses are no longer valid. The NFL's new data cannot tell us where exactly to look for insight, but it will allow us to both create new stories and to make old ones more complete.

\bibliographystyle{DeGruyter}
\bibliography{main}
\end{document}